\documentclass[a4paper,11pt]{article}

\usepackage{jcappub}
\usepackage{url}
\usepackage{bm}
\usepackage{amsmath}
\usepackage{appendix}

\RequirePackage{color}

\newcommand{\apj}{ApJ}
\newcommand{\mnras}{MNRAS}

\title{On the first crossing distributions in fractional Brownian motion and the mass function of dark matter haloes}

\author[a]{Nicos Hiotelis,}
\author[b,c,d]{Antonino Del Popolo}

\affiliation[a]{1st  Lyceum of Athens, Ipitou 15, Plaka, 10557,
Athens,  Greece}
\affiliation[b]{Dipartimento di Fisica e Astronomia, University Of Catania, \\
Viale Andrea Doria 6, 95125, Catania, Italy}
\affiliation[c]{INFN sezione di Catania,\\
Via S. Sofia 64, I-95123 Catania, Italy}
\affiliation[d]{International Institute of Physics, Universidade Federal do Rio Grande do Norte, \\
59012-970 Natal, Brazil}

\emailAdd{adelpopolo@oact.inaf.it}
\emailAdd{hiotelis@ipta.demokritos.gr}

\abstract{We construct  an integral equation for the first crossing distributions for fractional Brownian motion in the case of a constant barrier and we present an exact analytical solution. Additionally we present first crossing distributions derived by simulating paths from fractional Brownian motion.  We compare the results of the analytical  solutions with both those of simulations and those of some approximated solutions which have been used in the literature. Finally, we present  multiplicity functions for dark matter structures  resulting from our analytical approach and we compare with those resulting  from N-body simulations.\\
   We show that the results of analytical solutions are in good agreement with those of path  simulations but differ significantly from those derived from approximated solutions. Additionally, multiplicity functions derived from fractional Brownian motion are poor fits of the those which result from N-body simulations. We also present comparisons with other models which exist in the literature and we discuss different ways of improving the agreement between analytical results and N-body simulations.
}


\begin{document}

\maketitle

\flushbottom

\section{Introduction}
 The picture of structure formation in the Universe by the process of gravitational instability can be approximated analytically  by the excursion set model. This model is  based on the ideas of  \citet{prsc74} and on their extensions which are presented in the pioneered works of  \citet{pea}, \citet{boet91} and \citet{laco93}. In fact, the problem of structure formation is reduced to a first crossing problem, a problem that is present in many branches of science, \cite{red} . Formally, this problem leads to the construction of integral equations and in some cases, their analytical solutions can be found using integral transforms. Such an approach we present in this paper.
  An analytical solution for the first crossing distribution of a constant barrier, when  the initial density field obeys a fractional Brownian motion is derived. Additionally first crossing distribution are calculated numerically from simulations of the paths of random walks. We compare our results first with an approximated analytical formula used in the literature  and second with the results of N-body simulations.\\
 In Sect.2 we describe the physical picture of the EPS methods and we construct  the integral equation for the first crossing distribution . In Sect.3 the integral equation is solved in terms of Mellin transforms and the analytical expression for the first crossing distribution is found. In Sect.4 we describe the method of path simulations. In  Sect.5 we present our results  that additionally contain multiplicity functions which are compared with the results of N-body simulations. We also compare our  results to those of the model of \cite{ms}. Finally in Sect.6 a short discussion is given.

 \section{General Integral Equation for the first crossing distribution}
 Let a initial snapshot of the Universe.
   We find the density perturbation around a point of the Universe  using a filter $W$. This usually is assumed spherically symmetric and depends on a characteristic radius $R$. Then, the smoothed density perturbation  at the center of the spherical region is $\delta(R)=\int W(r;R)\hat{\delta} (r)4\pi r^2\mathrm{d}r$ where $\hat{\delta} (r)$ is the density at distance $r$ from the center of the spherical region. Since the Universe is homogeneous at large scales the expected value of $\delta$ is zero when the radius $R$ of the filter tends to very large values. We then define a critical value of $\delta$ named barrier, $\delta_c$. Decreasing the radius of the filter the resulting values of $\delta=\delta(R)$ vary. So in the plane $(R,\delta)$ we have a random walk. Let that for some value of the  filter radius $R$, which is reached first, and obviously is the maximum one, named $R_{max}$, a value of $\delta$ equal to  $\delta_c$ is found. Then it is assumed that a structure of radius $R_{max}$ is formed around that point. This structure is of mass $M$ given by $M=4{\pi}\rho_b\int W(r,R_{max})r^2\mathrm{d}r$. Thus, the problem of the formation of structures becomes a first crossing problem.  Since in a hierarchical scenario the variance $S$ of the overdensity at scale $R$ is a decreasing function of $M$ that is contained in this region,  and $M$ is an increasing function of $R$ we can consider a random walk on the plane $(S,\delta)$ that starts from the point $(0,0)$  and evolves as $S$ increases ($R$ decreases). Note that the role of $S$ is completely analogous to the role of time $t$ in one-dimensional random walks on $x$-axis.\\
    The correlation of values of $\delta$ between scales is given by the autocorrelation function that is
  \begin{equation}
  \label{neq1}
  \langle\delta(R)\delta(R')\rangle=\frac{1}{2\pi^2}\int_{0}^{\infty}k^2P(k){\widehat{W}_f}(k;R){\widehat{W}_f}(k;R')\mathrm{d}k
  \end{equation}
  where $P$ is the power spectrum and $\widehat{W}$ is the Fourier transform of the filter.
   The nature of the walks on the $(S,\delta)$ plane depends on the form of the smoothing kernel.
  If the smoothing kernel is the k-sharp filter then the correlation between two different scales $S$ and $S'$  is given the relation
   \begin{equation}
   \label{neq2}
   \langle\delta(S)\delta(S')\rangle=\min(S,S')
   \end{equation}
   which leads to a Brownian motion. Thus, in this case random walks have no memory. But for  more realistic filters  (as for example a top-hat in real space or a Gaussian one ) the correlation between scales has a different form and the walks are not memory-less any more. In fact for a Gaussian density field,  using these two more realistic filters the correlation between scales is different to that given in Eq.\ref{neq2}, see for example \cite{magr},
 and thus a realistic approximation of the true nature of random walks on the $(S,\delta)$ plane, requires the study of stochastic processes with memory as for example the fBm which is studied here.\

 Let $f(S)$ the first crossing distribution in the walks described above. Thus $f(S)\mathrm{d}S$ is the probability a walk passes from the first time the barrier $\delta_c$ at between $S, S+\mathrm{d}S$. Consequently $P_1\equiv\int_{0}^Sf(S')\mathrm{d}S'$ equals to the probability the walk has  crossed the barrier at some value less than $S$. We define as $P(\delta,S)$ the probability of finding a walk at $(\delta,S)$ that has never crossed the barrier for values $<S$.  The following equation holds:
 \begin{equation}
  \label{a1}
 \int_{-\infty}^{\delta_c}P(\delta,S)\mathrm{d}\delta+\int_{0}^Sf(S')\mathrm{d}S'=1
 \end{equation}
  Differentiating with respect to $S$  we have:
 \begin{equation}
  \label{a2}
 f(S)=-\int_{-\infty}^{\delta_c}\frac{\partial}{\partial S} P(\delta,S)\mathrm{d}\delta
 \end{equation}
 If we denote by $P_0(\delta,S)$ the probability of the random walk is at $(S,\delta)$,
 then,
 \begin{equation}
  \label{a3}
 P(\delta,S)=P_0(\delta,S)-\int_{0}^{S}f(S')P_0[\delta,S/S_c=S']\mathrm{d}S'
 \end{equation}
 where $P_0[\delta,S/Sc=S']$ is the probability the walk is at $(S,\delta)$ given that it has passed from the barrier for first time at at $S'$.\\ If the walk is memoryless, as for example in the case of Brownian motion, one can make use of the strong Markov property and replace $P_0[\delta,S/Sc=S']$ by $P_0[\delta,S/\delta_c,S']$ (\cite{zhang}.  This cannot be done in  the case of the  fractional Brownian motion  (fBm) since it  is a motion with memory. Thus, a different approach has to be used. First, we recall that a fBm with Hurst exponent $H$ has
 \begin{equation}
  \label{ab4}
 P_0(\delta,S)=\frac{1}{\sqrt{2\pi S^{2H}}}
 e^{-\frac{{\delta}^2}{2S^{2H}}}
 \end{equation}
 and  propagator,
 \begin{equation}
  \label{ab4a}
 P_0(\delta,S/\delta_0,S_0)=\frac{1}{\sqrt{2\pi (S-S_0)^{2H}}}
 e^{-\frac{(\delta-\delta_0)^2}{2(S-S_0)^{2H}}}
 \end{equation}
 Hurst exponent is a parameter that affects the behavior of random walks and its role is clarified in section 4. \\
  The integral equation for $f$ can be written as,
  \begin{equation}
  \label{a5}
 P_0(\delta_c,S)=\int_{0}^{S}f(S')P_0(\delta_c,S/\delta_c,S')\mathrm{d}S'
 \end{equation}
  Using the  expressions of Eq.\ref{ab4} and Eq.\ref{ab4a} we have,
  \begin{equation}
  \label{a6}
 e^{-\frac{C}{S^{2H}}}=\int_{0}^{S}f(S')\left(\frac{S}{S'}-1\right)^{-H}\left(\frac{S}{S'}\right)^H\mathrm{d}S'
 \end{equation}
 where $C\equiv\frac{{\delta_c}^2}{2}$. With the use of Heaviside function $\Theta$ Eq.\ref{a6} can be written as,
 \begin{equation}
  \label{a7}
 e^{-\frac{C}{S^{2H}}}=\int_{0}^{\infty}f(S')\left(\frac{S}{S'}-1\right)^{-H}\left(\frac{S}{S'}\right)^H \Theta\left(\frac{S}{S'}-1\right)\mathrm{d}S'
 \end{equation}
 or in a more convenient form as,
 \begin{equation}
 \label{a8}
 e^{-\frac{C}{S^{2H}}}=\int_{0}^{\infty}y(S')g\left(\frac{S}{S'}\right)\frac{\mathrm{d}S'}{S'}
 \end{equation}
 where $y(S')=S'f(S')$ and $g\left(\frac{S}{S'}\right)=\left(\frac{S}{S'}-1\right)^{-H}\left(\frac{S}{S'}\right)^H \Theta\left(\frac{S}{S'}-1\right)$.
 In Sect.3 we deal with the analytical solution of Eq. \ref{a8}. We will show  that the solution of Eq. \eqref{a5} is a very good approximation to the first crossing distribution of fBm and it improves significantly analytical expressions which have been used in the literature so far, \textbf{\cite{pan}, \cite{panetal}}.\\

  \section{Analytical solutions}
  In the appendix we present the analytical solution of Eq.\ref{a8}. This solution can be approximated by a simpler expression
  which,  as we will show below,  works very satisfactory, as follows:
  Changing the variable in Eq. \ref{a24} to $x=(1+t)^{2H}$ we can write,
  \begin{equation}
  \label{a24a}
  f(S)= \frac{\sin(\pi H)}{2\pi}{\nu}^2S^{-2H}\int_{1}^{\infty}[x^{\frac{1}{2H}}-1]^{H-1}e^{-\frac{1}{2}{\nu}^2S^{1-2H}x}\mathrm{d}x
  \end{equation}
  where $\nu=\delta_c/\sqrt{S}$.\\
  Since the main contribution to the integral comes from the values of $x$ close to unity we can use $x^{\frac{1}{2H}}-1\approx \frac{x-1}{2H}$. Then, the above equation is written as,
   \begin{equation}
  \label{a24b}
  f(S)= \frac{\Gamma(H)\sin(\pi H)}{\pi H^{H-1}}\frac{{\delta_c}^{2(1-H)}}{S^{1+2H(1-H)}}e^{-\frac{1}{2}\frac{{\delta_c}^2}{S^{2H}}}
  \end{equation}
  It is a simple exercise to check that for $H=1/2$ the above equations give the well known Markov density (or otherwise inverse Gaussian),
  \begin{equation}
  \label{a25}
  f_{ing}(S)= \frac{\delta_c}{\sqrt{2\pi}}S^{-\frac{3}{2}}e^{-\frac{\delta_c^2}{2S}}
  \end{equation}
  In Fig.1. we compare the results from equations \eqref{a24} and \eqref{a24b}. These expressions coincide for $H=1/2$ but, as it is shown in this figure,  they are very close for other values of $H$  in the range [0.4,0.6], which is interesting in a variety of physical problems.\\
   We note that if the assumption $P_0[\delta,S/Sc=S']=P_0[\delta,S/\delta_c,S']$ discussed in Sect.2, is applied in the fBm case then the integral equation that results from Eq.\ref{a3} has a solution
    \begin{equation}
     \label{a26}
     f(S)=\frac{2H\delta_c}{\sqrt{2\pi}S^{H+1}}e^{-\frac{\delta^2_c}{2S^{2H}}}
     \end{equation}
     (see for example \cite{pan},\cite{panetal}, \cite{hiot1}).
   Since the above assumption is not correct for the fBM, the result of  Eq.\ref{a26} can be considered  only as an approximate solution. Its validity will be checked in the following.\\
   \cite {ms} in their study of walks with correlated steps give for the first crossing distribution  in the case of a constant barrier, the following approximation,
   \begin{equation}
     \label{a27}
     f_{M-S}(S)=f_{ing}R(\Gamma, \nu)
     \end{equation}
     where
    \begin{equation}
     \label{a28}
     R(\Gamma, \nu)=\frac{1}{2}\left[\frac{1+\mathrm{erf}(\Gamma\nu/\sqrt{2})}{2}+\frac{e^{-\frac{1}{2}{\Gamma}^2{\nu}^2}}{\sqrt{2\pi}\Gamma\nu}\right]
     \end{equation}
     In the above relation  $\mathrm{erf}$ is the error function and $\Gamma$ depends on the power spectrum and the kernel used to smooth the density field. In fact $\Gamma^{-1}$ acts as the $\nu$ scale below which the correction to the completely correlated distribution becomes important. Thus for $\Gamma^{-1} \rightarrow 0$ the result of Eq. \eqref{a27} tends to the complete correlated case ( see Eq.1 in \cite{ms}). Increasing values of  $\Gamma^{-1}$ increases the correction to the complete correlated case.

  \section{Simulations for fBM}
  Fractional Brownian motion $B^H$, can be represented as an integral of a proper deterministic kernel with respect to an ordinary Brownian motion.  If $t$   denotes the time then we can write ,
  \begin{equation}
  \label{b1}
  B^H(t)= \int_0^tK_H(t,u)\mathrm{d}W(u)
  \end{equation}
   where $K_H$ is the deterministic kernel and $H\in (0,1)$. $W$ denotes the Brownian motion.\\
  A kernel that predicts the fractional Brownian motion is the  Molchan-Golosov kernel (\cite{molgol}). For $H>1/2$ it is given by
    \begin{equation}
   \label{b8}
   K_H(t,u)=\left(H-\frac{1}{2}\right)c_Hu^{\frac{1}{2}-H}\int_{u}^ty^{H-\frac{1}{2}}(y-u)^{H-\frac{3}{2}}\mathrm{d}y,~~0<u<t
   \end{equation}
    and for $H\leq 1/2$
    \begin{eqnarray}
   \label{b9}
   K_H(t,u)=c_H\left(\frac{t}{u}\right)^{H-\frac{1}{2}}(t-u)^{H-\frac{1}{2}}\nonumber\\
   -\left(H-\frac{1}{2}\right)c_Hu^{\frac{1}{2}-H}\int_{u}^ty^{H-\frac{3}{2}}(y-u)^{H-\frac{1}{2}}\mathrm{d}y,~~0<u<t
   \end{eqnarray}
   The constant is given by,
   \begin{equation}
   \label{b10}
   c_H=\frac{1}{\Gamma\left(H+\frac{1}{2}\right)}\left[\frac{2H\Gamma\left(H+\frac{1}{2}\right)\Gamma\left(\frac{3}{2}-H\right)}{\Gamma(2-2H)}\right]^{\frac{1}{2}}
   \end{equation}
   The autocorrelation for fBm is given by
    \begin{equation}
   \label{egnb1}
   \langle B_H(t)B_H(t')\rangle=\frac{1}{2}[t^{2H}+{t'}^{2H}-|t-t'|^{2H}]
   \end{equation}
  \begin{figure}
      \includegraphics[width=12cm]{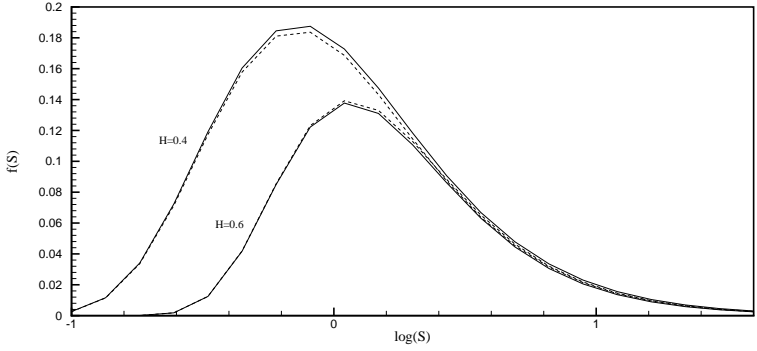}
      \caption{ A comparison between the predictions of Eqs \ref{a24} and \ref{a24b}. We note that for $H=1/2$ the results coincide, but as we see in this figure they are very close at least for values of $H$ around $1/2$. \emph{Solid} lines are the predictions of  Eq. \ref{a24} while \emph{dashed} lines  are those of Eq.\ref{a24b}.}\label{fig1}
     \end{figure}

  We  split the integral of Eq.\ref{b1} into a sum following the \cite{dela} method. We use  $n$ time intervals of lengths $\Delta t_i=t_i-t_{i-1}$. Thus we write,
  \begin{equation}
  \label{b2}
  B^H(t)= \sum_{i=1}^n\int_{t_{i-1}}^{t_i}K_H(t,u)\mathrm{d}W(u)=\sum_{i=1}^n\int_{t_{i-1}}^{t_i}K_H(t,u)\frac{\mathrm{d}W(u)}{\mathrm{d}u}\mathrm{d}u
  \end{equation}
  where $t_0=0$ and $t_n=t$.\\
  \begin{figure}
      \includegraphics[width=12cm]{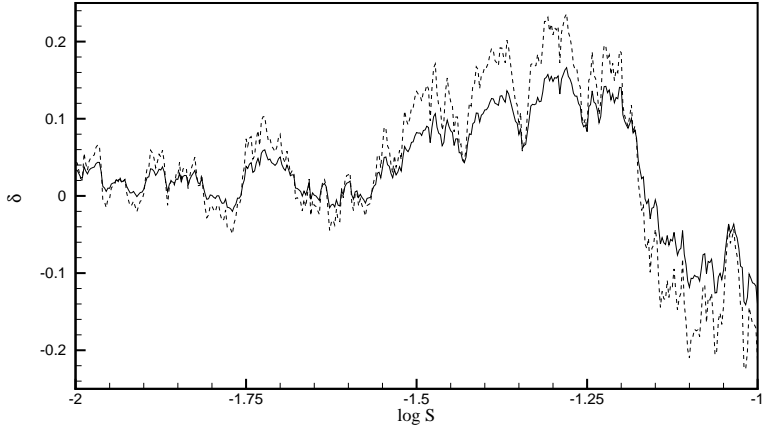}
      \caption{ A detail from two random walks. The \emph{dashed} line  corresponds to a walk predicted by a fBm with $H=0.45$ and the \emph{solid}  line to a walk with $H=0.55$. Both have been predicted from the same Brownian vector and thus the role of the deterministic kernel is clear.}\label{fig2}
     \end{figure}
     \begin{figure}
      \includegraphics[width=12cm]{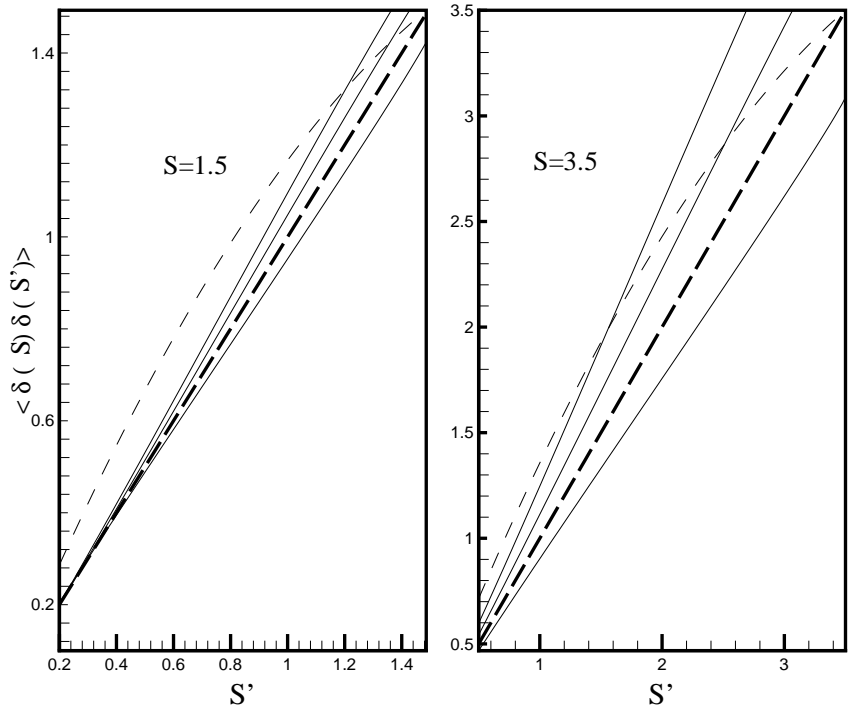}
      \caption{ A comparison of correlation between scales is shown. The left snapshot shows the correlation between the scale of mass
       $ M=1.18\times 10^{13}M_{\odot}h^{-1}$, which corresponds to $S=1.5$, with scales of larger mass. The largest  mass, $M'=1.22\times 10^{15}M_{\odot}h^{-1}$, corresponds to $S'=0.2$. From bottom to top, solid lines are the predictions of Eq.\ref{egnb1} of  fBm for $H=0.45$, $H=0.55$ and $H=0.6$ respectively. The thin dashed line shows the predictions of Eq.\ref{neq1} for a top-hat, in real space filter, and the thick dashed line shows the Brownian case. The right snapshot is similar but for $ M=7.9\times 10^{11}M_{\odot}h^{-1}$, which corresponds to $S=3.5$ and $M'=1.89\times 10^{14}M_{\odot}h^{-1}$, which corresponds to $S'=0.5$.}\label{fig3}
      \end{figure}

     \begin{figure}
      \includegraphics[width=12cm]{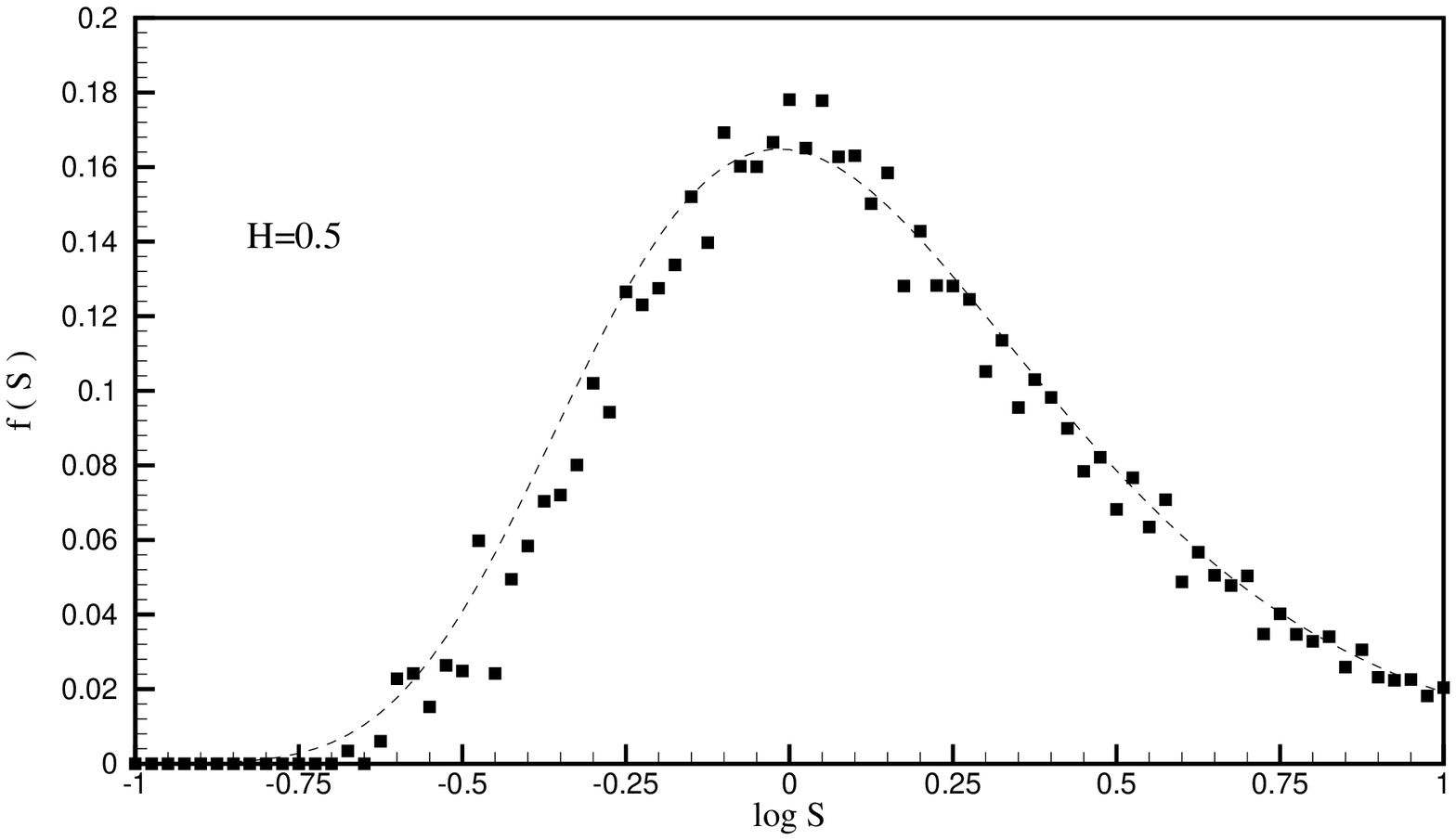}
      \caption{ Monte Carlo simulations vs analytical results for Brownian motion (H=1/2). The\emph{ dashed} line  corresponds to the predictions of formulae Eqs \ref{a24}, \ref{a25} and \ref{a26}, which coincide. \emph{Squares} are the results of path simulations. }\label{fig4}
      \end{figure}
  We assume that the quantity $\frac{\mathrm{d}W(u)}{\mathrm{d}u}$ is constant in the time interval $[t_{i-1},t_i]$ and equals to $\Delta W_i/\Delta t_i$. Thus for $t=t_j$ we can write,
  \begin{equation}
  \label{b3}
  B^H(t_j)= \sum_{i=1}^j\frac{\Delta W_i}{\Delta t_i}\int_{t_{i-1}}^{t_i}K_H(t_j,u)\mathrm{d}u
  \end{equation}
  \begin{figure}
      \includegraphics[width=12cm]{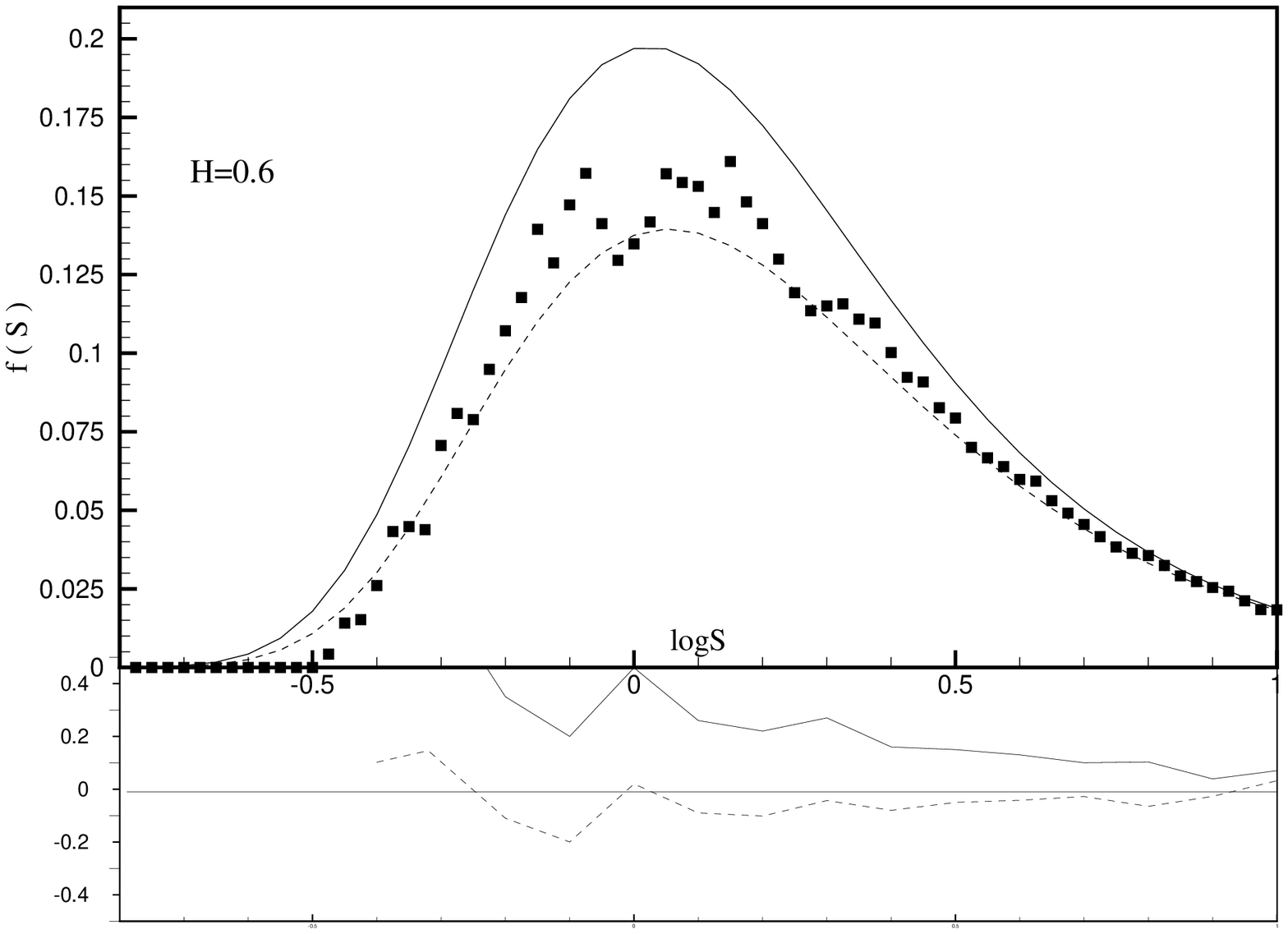}
      \caption{ Monte Carlo simulations vs analytical results for fractional  Brownian motion (H=0.6). The \emph{dashed}  line  corresponds to the predictions of formula Eqs \ref{a24} and the \emph{solid} line to the approximate formula given in Eq. \ref{a26}. \emph{Squares} are the results of path simulations of Eq.\ref{b10a}. The lower panel shows the relative error $(f_{analyt}-f_{monte carlo})/f_{monte carlo}$ for the above mentioned formulae.}\label{fig5}
      \end{figure}
  The steps to be followed are:\\
  First we divide the interval $[0,t]$ into $n$ intervals. Second we calculate $C_{i,j}=\frac{ \int_{t_{i-1}}^{t_i}K_H(t_j,u)\mathrm{d}u}{\Delta t_i}$ for $j=1,2...n$. Then we choose the numbers $\Delta W_i, i=1,2...n$ from a distribution,
   \begin{figure}
      \includegraphics[width=12cm]{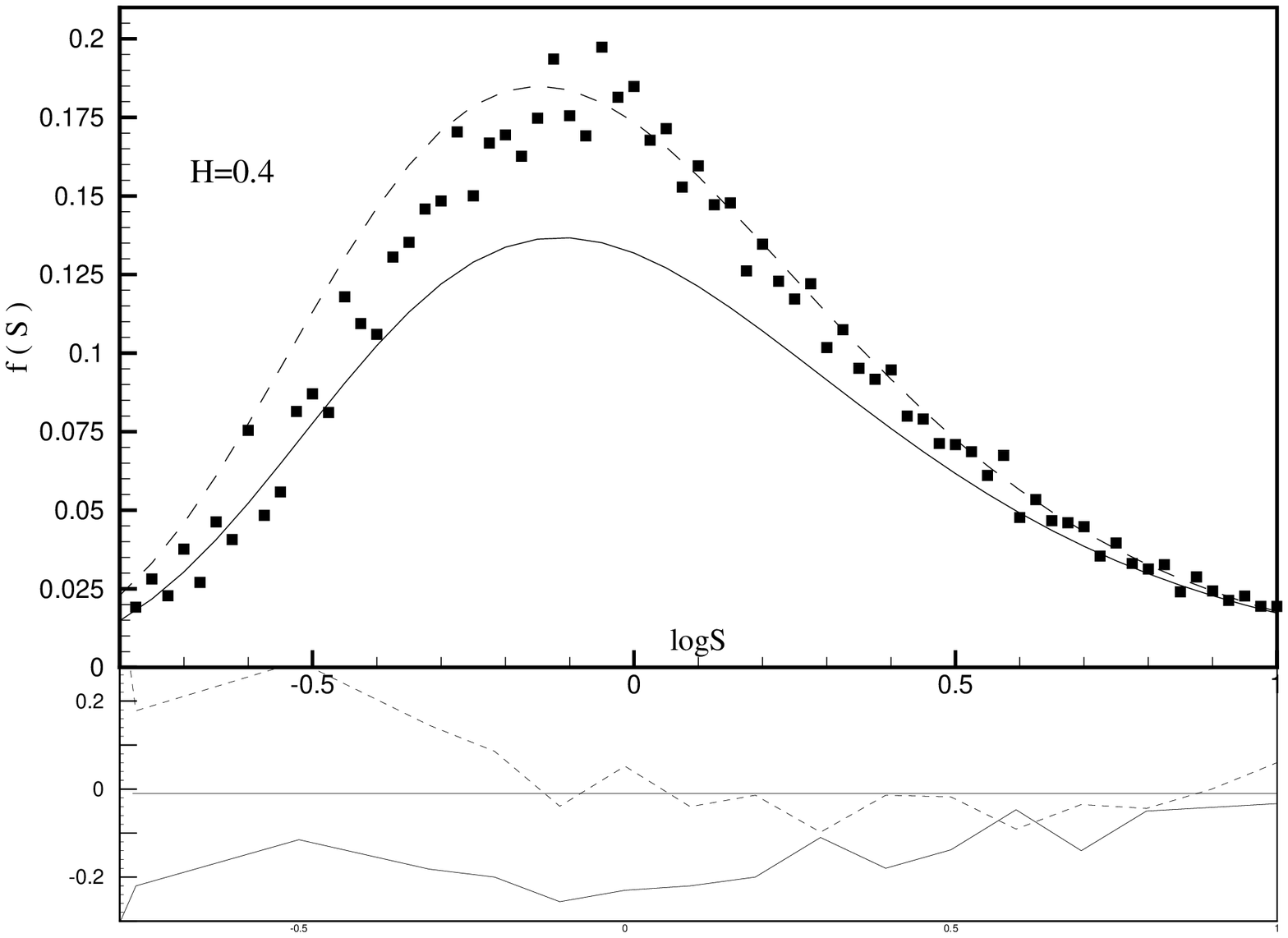}
      \caption{ Monte Carlo simulations vs analytical results for fractional Brownian motion (H=0.4). The\emph{ dashed}  line  corresponds to the predictions of formula Eqs \ref{a24} and the \emph{solid} line to the approximate formula given in Eq. \ref{a26}. \emph{Squares} are the results of path simulations of Eq.\ref{b10a}. The lower panel shows the relative error $(f_{analyt}-f_{monte carlo})/f_{monte carlo}$ for the above mentioned formulae.} \label{fig6}
      \end{figure}

  \begin{equation}
  \label{b4}
  P(\Delta W_i)=\frac{1}{\sqrt{2\pi \Delta t_i}}e^{-\frac{(\Delta W_i)^2}{2(\Delta t_i)}}
  \end{equation}
  that is a Gaussian with zero mean value and $\sigma^2_i=\Delta t_i$. This can be done by choosing values $X_1,X_2,..X_N$ from a normal $N(0,1)$ distribution and then define $\Delta W_i=X_i\sqrt{\Delta t_i}$.\\
  Then, for $j=1,2,...n$ we calculate the quantity
  \begin{equation}
  \label{b5}
  B^H(t_j)= \sum_{i=1}^jC_{i,j}\Delta W_i
  \end{equation}

  Thus, for any set $\Delta W_i, i=1,2...n$ we have the positions of a tracer particle at times $t_1,t_2,...t_n$. For $N$ such sets we will have the positions of $N$ such traces particles that can be used for our analysis. Note that for any value $j$ we have
  \begin{equation}
  \label{b6}
  \langle B^H(t_j)\rangle= 0
  \end{equation}

  since $B^H(t_j)$ is a sum of Gaussians with zero mean. Additionally,
  \begin{equation}
  \label{b7}
   Var[B^H(t_j)]=\sum_{i=1}^{j}C^2_{i,j}
   \end{equation}

     \begin{figure}
      \includegraphics[width=12cm]{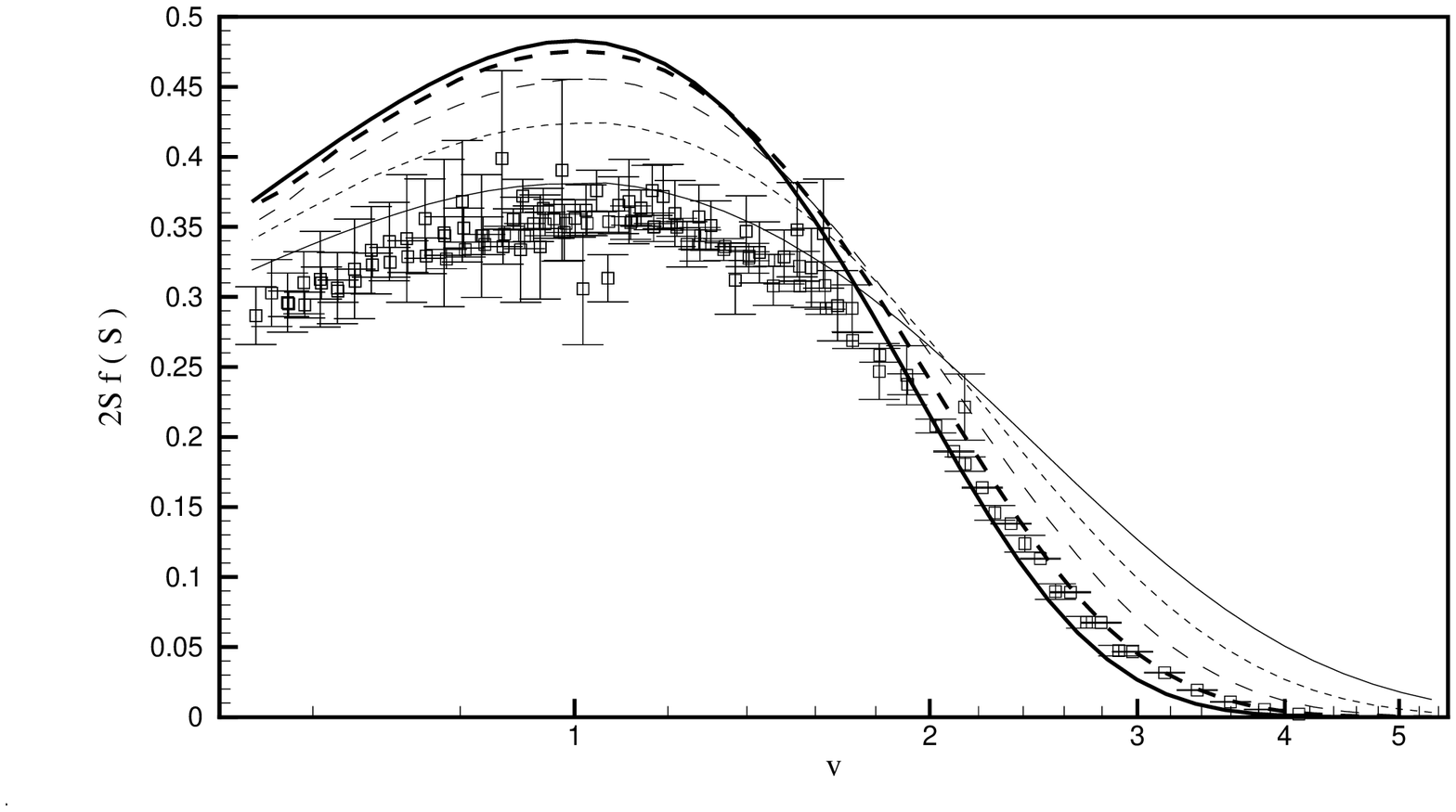}
      \caption{ A sequence of multiplicity functions for various values of $H$, where the  analytical solution for $f(S)$ given by Eq. \ref{a24} is used. \emph{Thick solid} line was predicted for $H=0.5$, line with \emph{thick dashes} for $H=0.45$, \emph{dashed line} for $H=0.40$, line with \emph{small dashes} for $H=0.35$ and the \emph{solid} line for $H=0.30$. \emph{Squares}  are the results of N-body simulations of Tinker et al, \cite{tin}. }\label{fig7}
      \end{figure}
     \begin{figure}
      \includegraphics[width=12cm]{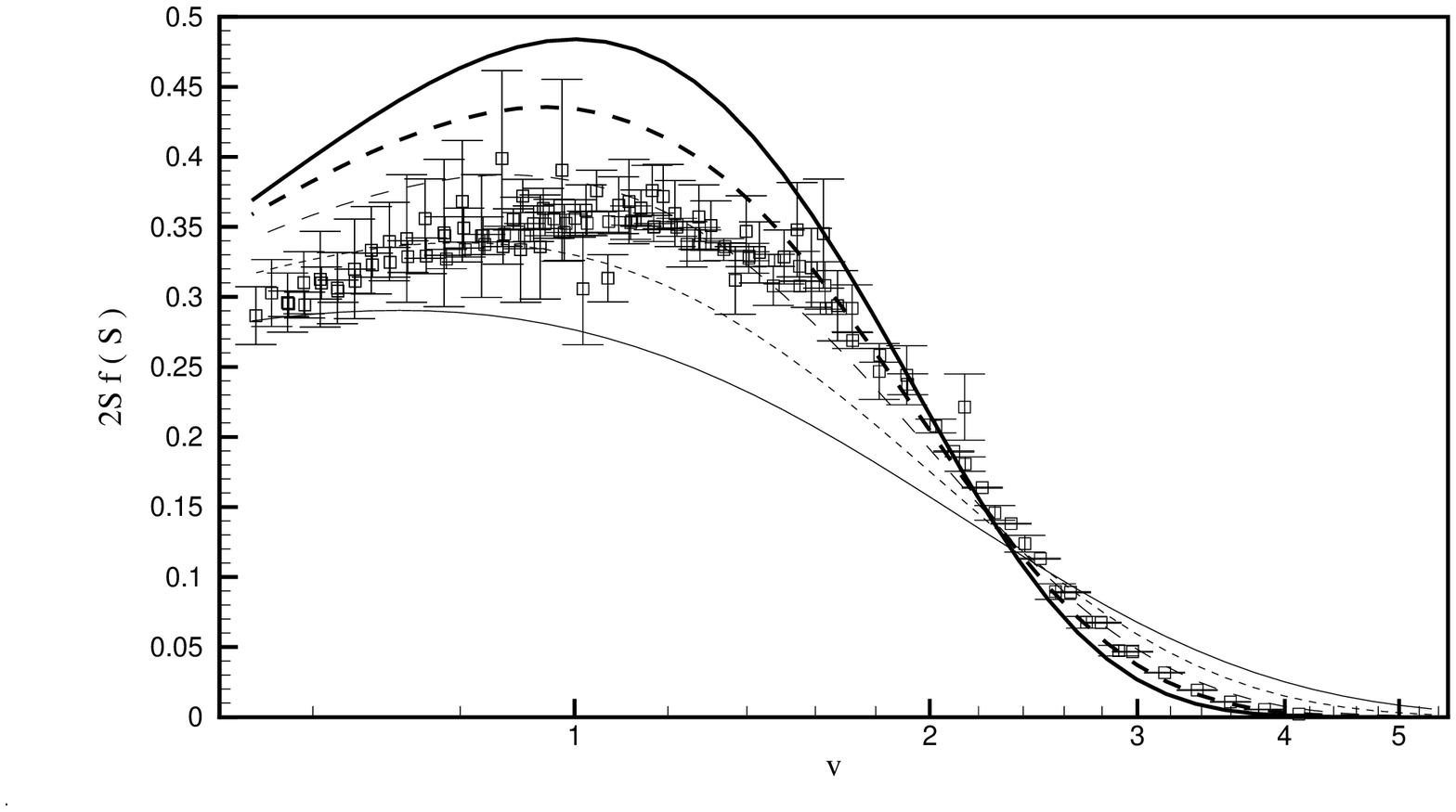}
      \caption{ A sequence of multiplicity functions for various values of $H$, where the approximate solution for the first crossing distribution $f_{approx} (S)$ given by Eq. \ref{a26} is used. \emph{Thick solid} line was predicted for $H=0.5$, line with \emph{thick dashes} for $H=0.45$, \emph{dashed line} for $H=0.40$, line with \emph{small dashes} for $H=0.35$ and the \emph{solid line} for $H=0.30$. \emph{Squares} are the results of N-body simulations of Tinker et al \cite{tin}. }\label{fig8}
      \end{figure}

         Obviously, for the problem we face here $t$ must be replaced by $S$ and $B^H(t)$ by $\delta(S)$. For a  kernel which describes properly the fBm,  the sum in Eq.\ref{b7}  should be equal to $S_j^{2H}$ as it is known from Eq.\ref{ab4}, and this was used as  a test for the reliability  of our simulation.\\
     We divided the interval $[S_{min},S_{max}]$ to $N$ intervals and we constructed the paths for $N_{paths}$ tracer particles. Typically we used $S_{min}=10^{-4},S_{max}=10$, $N=2000$ and $N_{paths}=50000$. Note that a large number of $N$ is essential for the accuracy of the integration. The first crossing time of every path is found by checking at every of its $N$ steps if it passes the barrier. We then define a grid of $N_G$ intervals of the form $[D_{i-1},D_i]$ and the number of paths which have their  first crossing times in this interval is  found, let $N_i$. Finally we calculate the first crossing distribution at $d_i=(D_{i-1}+D_i)/2$ by
     \begin{equation}
     \label{b10a}
     f_{sim}(d_i)=\frac{N_i}{N_{paths}(D_{i}-D_{i-1})}
     \end{equation}
     Typically we use $N_G=200$.  In both grids ($N$ and $N_G$) we used a spacing constant in $\log S$.\\
     In Fig.2 we present two walks.  One  for $H=0.45$ and the other  for $H=0.55 $. Since from Eq.\ref{ab4a} is known that the expectation value of $[\delta(S)-\delta(S')]^2$ is $E[\delta(S)-\delta(S')]^2=\mid S-S'\mid^{2H}$ we have for the correlation $\mathrm{Cor}$ the following relation,
     \begin{equation}
     \label{b11}
     \mathrm{Cor}(\Delta S)\equiv E\left[(\delta(S)-\delta(S-\Delta S))(\delta(S+\Delta S)-\delta(S))\right] =\mid\Delta S\mid^{2H}[ 2^{2H-1}-1]
     \end{equation}
     This shows that the correlation of values of $\delta$ can be either positive, for $H>1/2$ or negative for $H<1/2$. This is reflected to the walks of Fig.2. A walk predicted for $H>1/2$ is  a persisting case. It persists to its history up to the current "time". Thus it appears smoother than a walk with $(H<1/2)$ which  is anti-persisting. This last,  has the trend to change its direction. This makes it very noisy. Note that, the above two walks are predicted using the same Brownian vector $\Delta W_1,\Delta W_2,...\Delta W_ N$ and thus the differences reflect the role of the deterministic coefficients $C_{ij}$.

      \section{First crossing distributions and multiplicity functions }
     There is a question which naturally arises before any attempt to apply the results of the above analysis to the problem of structure formation in the Universe: What are the autocorrelation functions between scales which result from equations \ref{neq1} and \ref{egnb1} respectively ? Is for example the fBm approach more close to the physical problem  than the approach of the memoryless Brownian motion? In order to answer to such  questions we used  Eq. \ref{neq1}, a power spectrum which is described below,  a top-hat, in real space, filter  and we constructed  the autocorrelation function which is presented in Fig.3. We compare the above prediction with the results of fBm using   Eq.\ref{egnb1}. It is shown in  Fig. 3 that  a fBm with $0.55\leq H \leq 0.6$ approximates the correlation  better than the memoryless model. \textbf{We note that memory effects are included in the problem of structure formation by \cite{magr}, where the correlation between scales,as it results from Eq. \ref{neq1},  is approximated by an analytical formula and then a path integral method is used in order to approximate  first  crossing distributions. Obviously, our study  is an alternative approach to the same problem.} \\
          In our calculations we used the barrier $\delta_c(z)=1.686/D(z)$ where $D(z)$ is the growth factor derived by the linear theory, normalized to unity at the present epoch. This is a constant  barrier since it does not  depend on $S$. The physical picture is that in an Einstein-de Sitter Universe, a spherical region collapses at $z$, if the linear extrapolation of its initial value $\delta_{in}$ up to the present epoch equals to $\delta_c (z)$ (see for example \cite{peeb80}).\\
     \begin{figure}
      \includegraphics[width=12cm]{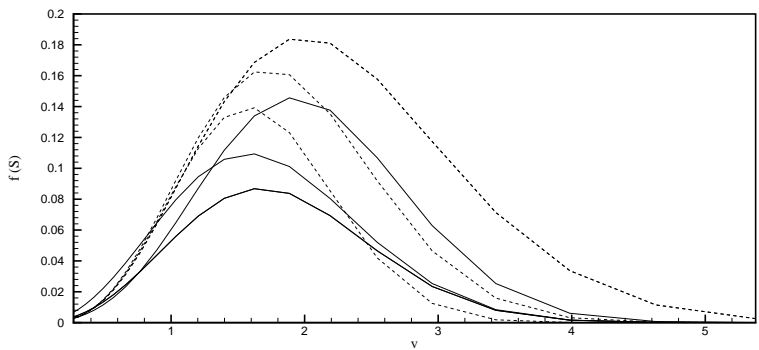}
      \caption{ First crossing distributions f(S) derived from equations  Eqs \ref{a24} and \ref{a27}. \emph{Dashed} lines are, from top to bottom, are the predictions of Eq. \ref{a24} for $H=0.4$, $H=0.5$ and $H=0.6$ respectively. Solid lines are the predictions of Eq.\ref{a27} for $\Gamma=1/4, \Gamma=1/3$ and $\Gamma=2/3$ respectively.}\label{fig9}
     \end{figure}
      \begin{figure}
      \includegraphics[width=12cm]{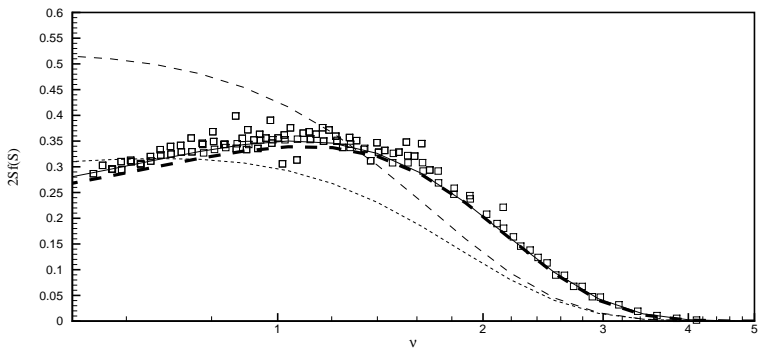}
      \caption{ A comparison of multiplicity functions predicted by Eq.\ref{a27} with the results of N-body simulations. \emph{Squares} are the results of \cite{tin} and the \emph{thin } solid line is the fit to the results given by their Eq.3. \emph{Large Dashes} and \emph{small dashes }  are the results of Eq. \ref{a27} for $\Gamma=1/3 $ and $\Gamma =2/3$ respectively. The \emph{thick} dashed line , that is an excellent fit to the results of N-body simulations of Tinker et al \cite{tin}, was predicted using $\Gamma^{-1}=1+\nu^{2.5}$. }\label{fig10}
     \end{figure}
     Our results were derived at $z=0$ using a flat model for the Universe with
  present day density parameters $\Omega_{m,0}=0.3$ and
   $ \Omega_{\Lambda,0}\equiv \Lambda/3H_0^2=0.7$ where
  $\Lambda$ is the cosmological constant and $H_0$ is the present day value of Hubble's
  constant. We have used the value $H_0=100~\mathrm{hKMs^{-1}Mpc^{-1}}$
  and a system of units with $m_{unit}=10^{12}M_{\odot}h^{-1}$,
  $r_{unit}=1h^{-1}\mathrm{Mpc}$ and a gravitational constant $ G=1$. At this system of units
  $H_0/H_{unit}=1.5276.$ Regarding  the power spectrum, we  employed the $\Lambda CDM$ formula proposed by
  \citet{smet98}.
     We present first crossing distributions which are derived by three different ways. The first way uses Eq.\ref{a24}, the second uses Eq. \ref{a26} and the third uses  Eq.\ref{b10a}. We name first crossing distributions  $f_{anal},f_{aprox}$ and $f_{sim}$ respectively.\\
     In  Fig.4 we present $f_{anal}$,and $ f_{sim}$ for $H=1/2$. We note that for $H=1/2$ the predictions of $f_{anal}$ and $f_{aprox}$ coincide. Thus this figure provides a check for the ability of simulations to predict results close to analytical ones. The agreement is satisfactory.\\
     In Figs 5 and 6 we present results for first crossing distributions for $H=0.60$ and $H=0.40$ respectively. In both figures solid lines show the predictions of the approximated first crossing distribution given by Eq.\ref{a26}. Dashed lines correspond to the predictions of the analytical solution Eq.\ref{a24}. Squares are the results of path simulations. It is clear that the results of Eqs \ref{a26} and \ref{a24} differ significantly. Thus, Eq.\ref{a26} cannot be considered as a good approximation of the first crossing distribution in fBm. As reqards the results of simulations can be considered as satisfactory. The noisy appearance can be improved using a larger number of paths and  a small number of grid points $N_G$. In any case it is clear that $f_{sim}$ is close to the results of Eq.\ref{a24}.\\
     In Figs 7 and 8 we compare  multiplicity functions, $2S f(S)$, derived by our models with the  results from N-body simulations which kindly  became available by J.Tinker (\cite{tin}). We have to note here  that open squares, that represent the results of N-body simulations,  come from  counting the number $N(M,z)$ of haloes of mass $M$ present in the simulation at $z$. A comparison should be correct if the quantities  $Sf(S)\mathrm{dln}S$ and $M^2\frac{N(M,z)}{\rho_b(z)}{\mathrm{dln}M}$ were equal. This ansatz is incorrect:  Whereas f(S) is a statement obtained by averaging over all walks in a Gaussian field, halos only form around special positions, for which the statistics are modified \cite{smt},\cite{para}, \cite{aetal}. Recent works have shown how to improve on this ansatz, by explicitly averaging only over special positions  \cite{coa}. However this problem may contributes significantly to the differences shown in Figures 4 and 5 but its study is beyond the scope of this paper. \\
        In Fig.7 a sequence of multiplicity functions for various values of $H$ are presented and are compared to the results of N-body simulations. All lines correspond to the exact analytical solution for $f(S)$ given by Eq. \ref{a24}. Thick solid line was predicted for $H=0.5$, line with thick dashes for $H=0.45$, dashed line for $H=0.40$, line with small dashes for $H=0.35$ and the solid line for $H=0.30$. Open squares are the results of N-body simulations, \cite{tin}. It is clear that the results of fBm are poor representations of the multiplicity functions derived by N-body simulations. Decreasing the value of $H$ the fit is improved for small haloes but it becomes worse  for heavier ones.\\
   A similar sequence is presented in Fig.8. Lines correspond to the same values of $H$ as in Fig.7 but for the approximated solution given by Eq. \ref{a26}. It is interesting that for $H\simeq 0.4$ the approximated solution is a good fit to the results of N-body simulations but obviously this cannot attributed to the fBM.\\
   Note that results for $H>1/2$ are more poor fits than those presented above and have no interest for the problem we face here.\\
   In Fig.9 we compare some first crossing distributions derived  from equations  Eqs \ref{a24} and \ref{a27}. Dashed lines are, from top to bottom, the predictions of Eq. \ref{a24} for $H=0.4$, $H=0.5$ and $H=0.6$ respectively. Solid lines are the predictions of Eq. \ref{a27} for $\Gamma=1/4, \Gamma=1/3$ and $\Gamma=2/3$ respectively. Although the values of $\Gamma$ and $H$ seem to be positively  correlated it is clear that no agreement can be reached.\\
   In Fig.10 a comparison of multiplicity functions predicted by Eq.\ref{a27} with the results of N-body simulations. Squares are the results of \cite{tin} and the thin  solid line is the fit to the results given by their Eq.3. Large dashes and small dashes  are the results of Eq. \ref{a27} for $\Gamma=1/3 $ and $\Gamma =2/3$ respectively. It is obvious that there is no constant value of $\Gamma$ that can fit the results of Eq.\ref{a27} with those of N-body simulations. Instead an excellent fit can be achieved for varying $\Gamma$. If we allow the correction to the complete correlated case to decrease with increasing halo mass then the fit is significantly improved. As an example we show the  thick dashed line , that is an excellent fit to the results of N-body simulations, which was predicted using $\Gamma^{-1}=1+\nu^{2.5}$. This is a form that best fits the data. Its physical meaning is that a fixed value of $\Gamma$ defined globally from the power spectrum and the filter is not appropriate.The scale below  which the correction to the completely correlated distribution becomes important ($\Gamma^{-1}$), should depend on the value of $S$. Expect for a few points of the noisy results of N-body simulations the relative error of this fit is better than a few percent.


 \section{Conclusions and Discussion}
  The construction of first crossing distributions is an interesting problem for many branches of science.In the first part of this paper we present a formula for fist crossing distribution in the case of fBm, that is a stochastic process with memory. What is the role of  a process with memory for the particular problem which is studied here? For example in the persisting case, a walk which has a history of rising values of its $\delta 's$, is more probable to continue rising and has a different probability to up-cross a certain barrier at a specific  larger $S$, than in the anti-persisting case, where it is preferable to change its direction. This difference redistributes first passage times and results to different mass functions. Additionally if a walk passes the barrier at $S'$ it has a different probability to cross a larger boundary  at $ S>S'$ for persisting and anti-persisting cases respectively. This results to a correlation between structures and could connect the characteristics of haloes with their environment but such a study is beyond of the scope of this paper.\\
   The formula presented is compared to path simulations of fBm and it is shown that it works very satisfactory and definitely much better than other approximated formulae which are exist in the literature \cite{pan},\cite{panetal},\cite{li}. In the second part  of this paper, fBm is used for the  problem of the formation of dark matter haloes in terms of the excursion set model. Various interesting approximations for this problem  that  refer mainly to the form of the barrier and  to the use of various filters, have been used. Additional studies are about  non-Gaussian initial density fields, diffuse barriers or even about models with anomalous diffusion (see for example  \cite{she98}, \cite{smt}, \cite{zhang}, \cite{coa}, \cite{magr}, \cite{hiop}, \cite{hiop1},  \cite{hiop2} and references therein). In this paper we used the simplest form of the barrier, that is a constant one,  but we assumed walks with memory which are governed by the fBm. We compared our predictions to those of \cite{ms} and we found that no constant value of $\Gamma$ can lead to an agreement between the resulting first crossing distributions. Additionally,  comparisons with the results of N-body simulations are  presented. We conclude that that fBm  is not able to approximate satisfactory these numerical results. The resulting curves that give the correct number of  small mass haloes overestimate the number of large mass haloes and those that  fit the number of large mass haloes overestimate the number of small mass haloes. We also presented a comparison between the predictions of  the formula of \cite{ms} and the results of N-body simulations. We found that there is a disagreement too that cannot be reduced for constant values of the parameter $\Gamma$.\\
  Before drawing any  final conclusions about the models studied in this paper  we must to take into account  once again, the remarks of the previous section regarding on the ansatz used for the comparison between the analytical results and those of N-body simulations. The improvement of this ansatz is necessary for reliable comparisons and consequently for the prediction of robust conclusions. However, in terms of the approach presented in this paper, we think that an attempt for the construction of a memory kernel which approximates better the correlation between various scales which results from the use of realistic filters, is an important problem. This problem is under study. \\
 \section{Acknowledgements}
\ We acknowledge  J.Tinker for kindly providing the results of their N-body simulations about multiplicity functions.

\section{Appendix}
Integral transforms are power tools for the solution of some integral equations as Eq.\ref{a8}. The solution that follows will be derived by the use of Mellin transforms. Thus, it it necessary to recall a few properties which are useful to proceed further. The Mellin transform of a function $g$ of the variable $s$ defined on $(0,\infty)$ is given by the relation:
   \begin{equation}
   \label{a9}
  M[g(s),p ]=\int_{0}^{\infty}s^{p-1}g(s)\mathrm{d}S=\hat{g}(p)
  \end{equation}
  Useful properties are,
  \begin{equation}
   \label{a9a}
  M[e^{-as},p ]=\frac{\Gamma(p)}{a^p}, a>0 ~~~M[g(as),p]=\frac{\hat{g}(p)}{a^p}, a>0
  \end{equation}
  \begin{equation}
   \label{a9b}
   M[s^ag(s),p]=\hat{g}(p+a),~~~M[g(s^a),p]=\frac{1}{\mid a \mid}\hat{g}\left(\frac{p}{a}\right),~a\neq 0
  \end{equation}
   where $\Gamma$ is the gamma function.\\
  Combining the above properties we can easily prove the following very useful, for our calculations,  relations
  \begin{equation}
    \label{a10}
  M[s^ae^{-ks^b},p ]=\frac{1}{\mid b \mid}\frac{\Gamma\left(\frac{p+a}{b}\right)}{k^{\frac{p+a}{b}}},~~k>0,~b\neq 0 ~~M\left[s^a g(s^b),p \right]=\frac{1}{\mid b \mid}\hat{g}\left(\frac{p+a}{b}\right),~b\neq 0
  \end{equation}
  Additionally,
  \begin{equation}
   \label{a11}
  M[W^{\beta}g(s),p]=\frac{\Gamma(p)}{\Gamma(p-\beta)}\hat{g}(p-\beta)
  \end{equation}
  where $W^{\beta}$ is the Weyl fractional derivative of order $\beta$, defined by the relation,
  \begin{equation}
  \label{a12}
  W^{\beta}g(s)=\frac{(-1)^n}{\Gamma(n-\beta)}\frac{{\mathrm d}^n}{\mathrm{d}s^n}\int_s^{\infty}(t-s)^{n-\beta-1}g(t)\mathrm{d}t
  \end{equation}
  where $\beta$ is positive and $n$ is the smallest integer greater than $\beta$ such as $n-\beta$ is greater than zero.\\
  We left last the property which will use first and justifies the way in which the integral equation is written. This is
  \begin{equation}
  \label{a13}
  M\left[\int_{0}^{\infty}y(s')g\left(\frac{s}{s'}\right)\frac{\mathrm{d}s'}{s'},p\right]=\hat{y}(p)\hat{g}(p)
  \end{equation}
  (see Eq. (8.3.18) in  \cite{intgr}).\\
  We Mellin transform Eq.\ref{a8} using Eq.\ref{a13} and we have
  \begin{equation}
  \label{a14}
  \frac{1}{2H}\Gamma\left(-\frac{p}{2H}\right)C^{\frac{p}{2H}}=\hat{y}(p)\hat{g}(p)
  \end{equation}
  where $\hat{y}(p)$ and $\hat{g}(p)$ are the Mellin transforms of $Sf(S)$ and $(S-1)^{-H}S^H\Theta(S-1)$ respectively.\\
  The calculation of $\hat{g}(p)$ is straightforward,
   \begin{eqnarray}
  \label{a15}
  \hat{g}(p)=\int_{0}^{\infty}S^{p-1}(S-1)^{-H}S^H\Theta(S-1)\mathrm{d}S=\int_{1}^{\infty}S^{p+H-1}(S-1)^{-H}\mathrm{d}S=\nonumber\\
  \int_{0}^{1}S^{-p-1}(1-S)^{-H}\mathrm{d}S=B(-p,-H+1)=\frac{\Gamma(-p)\Gamma(1-H)}{\Gamma(1-p-H)}
  \end{eqnarray}
  where $B$ is the beta function.\\
  Solving for $\hat{y}(p)$ we have,
 \begin{equation}
  \label{a16}
  \hat{y}(p)=M[Sf(S),p]=\hat{f}(p+1)=\frac{1}{2H\Gamma(1-H)}\frac{\Gamma\left(-\frac{p}{2H}\right)\Gamma(1-p-H)}{C^{-\frac{p}{2H}}\Gamma(-p)}
  \end{equation}
  Changing the variable $p\rightarrow 1-H-p$  we have,
  \begin{equation}
  \label{a17}
  \hat{f}(2-H-p)=\frac{1}{2H\Gamma(1-H)}\frac{\Gamma\left(\frac{p+H-1}{2H}\right)\Gamma(p)}{C^{\frac{p+H-1}{2H}}\Gamma(p+H-1)}
  \end{equation}
  We note that,
  \begin{equation}
  \label{a18}
  M[W^{1-H}\omega(S)]=\frac{\Gamma(p)}{\Gamma(p+H-1)}\hat{\omega}(p+H-1)
  \end{equation}
  Thus, if we find a function $\omega$ that satisfies,
  \begin{equation}
  \label{a19}
  \hat{\omega}(p+H-1)=
  \frac{1}{2H\Gamma(1-H)}\frac{\Gamma\left(\frac{p+H-1}{2H}\right)}{C^{\frac{p+H-1}{2H}}}
  \end{equation}
  then the right hand side of Eq.\ref{a17} will be $M[W^{1-H}\omega(S)]$. We are looking for  $\omega$ such as $\hat{\omega}(p)=\frac{1}{2H\Gamma(1-H)}\frac{\Gamma\left(\frac{p}{2H}\right)}{C^{\frac{p}{2H}}}$.
  Using the first property in  Eq.\ref{a10} we find that
  \begin{equation}
  \label{a20}
  \hat{\omega}(p)= \frac{1}{\Gamma(1-H)}M[e^{-CS^{2H}},p ]
  \end{equation}
  Thus finally,
  \begin{equation}
  \label{a21}
  \hat{f}(2-p-H)= \frac{1}{\Gamma(1-H)}M[W^{1-H}(e^{-CS^{2H}}),p ]
  \end{equation}
  From the second property of Eq.\ref{a10} we have that $\hat{f}(2-p-H)=M[S^{H-2}f(S^{-1}),p]$. Substituting in Eq.\ref{a21} we have
  \begin{equation}
  \label{a22}
  f(S^{-1})= \frac{1}{\Gamma(1-H)}S^{2-H}W^{1-H}(e^{-CS^{2H}})
  \end{equation}
  For $\beta=1-H$ in Eq.15 we have $n=1$ and the above equation is written as
  \begin{equation}
  \label{a23}
  f(S^{-1})= -\frac{1}{\Gamma(1-H)\Gamma(H)}S^{2-H}\frac{\mathrm{d}}{\mathrm{d}S}\int_{S}^{\infty}(t-S)^{H-1}e^{-Ct^{2H}}\mathrm{d}t
  \end{equation}
  Changing the variable of integration $t\rightarrow t+S$, performing the differentiation with respect to $S$ , substituting $S\rightarrow 1/S$ and changing once again the variable of the integration $t\rightarrow t/S$ we have the final form
  \begin{equation}
  \label{a24}
  f(S)= \frac{2HC\sin(\pi H)}{\pi}S^{-1-2H}\int_{0}^{\infty}t^{H-1}(1+t)^{2H-1}e^{-C\frac{(1+t)^{2H}}{S^{2H}}}\mathrm{d}t
  \end{equation}
  where we have used $\Gamma(H)\Gamma(1-H)=\frac{\pi}{\sin(\pi H)}$. \\

    \end{document}